\documentclass[aps,pre,twocolumn,superscriptaddress,nofootinbib]{revtex4-2}

\usepackage{amsmath}
\usepackage{graphicx}
\usepackage{amssymb}
\usepackage{color}
\usepackage{mathrsfs}
\usepackage{verbatim} 
\usepackage[colorlinks,linkcolor=blue,urlcolor=cyan,citecolor=blue,anchorcolor=blue]{hyperref}
\usepackage[T1]{fontenc}
\usepackage{orcidlink}

\begin{document}	

\title{Paradoxical non-Gaussian behavior in fractional Laplace motion with drift}

\author{Wei Wang\,\orcidlink{0000-0002-1786-3932}}
\affiliation{Institute of Physics \& Astronomy, University of Potsdam, 14476
Potsdam, Germany}
\author{Yingjie Liang\,\orcidlink{0000-0002-3201-6969}}
\affiliation{College of Mechanics and Engineering Science, Hohai University,
211100 Nanjing, China}
\affiliation{Institute of Physics \& Astronomy, University of Potsdam, 14476
Potsdam, Germany}
\author{Aleksei V. Chechkin\,\orcidlink{0000-0002-3803-1174}}
\affiliation{Institute of Physics \& Astronomy, University of Potsdam, 14476
Potsdam, Germany}
\affiliation{Faculty of Pure and Applied Mathematics, Hugo Steinhaus Center,
Wroc{\l}aw University of Science and Technology, 50-370 Wroc{\l}aw, Poland}
\affiliation{German-Ukrainian Core of Excellence, Max Planck Institute of
Microstructure Physics, Weinberg 2, 06120 Halle, Germany}
\affiliation{Asia Pacific Centre for Theoretical Physics, Pohang 37673, Republic
of Korea}
\author{Ralf Metzler\,\orcidlink{0000-0002-6013-7020}}
\email{rmetzler@uni-potsdam.de}
\affiliation{Institute of Physics \& Astronomy, University of Potsdam, 14476
Potsdam, Germany}
\affiliation{Asia Pacific Centre for Theoretical Physics, Pohang 37673, Republic
of Korea}

\begin{abstract}
We study fractional Laplace motion (FLM) obtained from subordination of fractional
Brownian motion to a gamma process, in the presence of an external drift that
acts on the composite process or of an internal drift acting solely on the
parental process. We derive the statistical properties of this FLM process and
find that the external drift does not influence the mean-squared displacement (MSD),
whereas the internal drift leads to normal diffusion, dominating at long times in
the subdiffusive Hurst exponent regime. We also investigate the intricate
properties of the probability density function (PDF), demonstrating that it
possesses a central Gaussian region, whose expansion in time is influenced by
FBM's Hurst exponent. Outside of this region the PDF follows a non-Gaussian
pattern. The kurtosis of this FLM process converges toward the Gaussian limit at
long times insensitive to the extreme non-Gaussian tails. Additionally, in the
presence of the external drift, the PDF remains symmetric and centered at $x=vt$.
In contrast, for the internal drift this symmetry is broken. The results of our
computer simulations are fully consistent with the theoretical predictions. The
FLM model is suitable for describing stochastic processes with a non-Gaussian
PDF and long-ranged correlations of the motion.
\end{abstract}

\maketitle

\section{Introduction}
\label{sec1}

The stochastic motion of a tracer particle is by-now routinely recorded by
single-particle tracking (SPT) experiments in various complex systems
\cite{shen17,manz15}, including biological cells, lipid membranes, polymer
solutions and porous media \cite{jaqa08,metz16,shin15,skau15}. The observed
motion typically deviates from the laws of regular Brownian motion (BM),
characterized by the linear time dependence of the mean-squared displacement
(MSD) $\langle x^2(t)\rangle\simeq t$ and the Gaussian probability density
function (PDF). Instead, anomalous diffusion is effected, characterized by a
power-law form $\langle x^2(t)\rangle\simeq t^\alpha$ of the MSD with the
anomalous diffusion exponent $\alpha$. Subdiffusion with $0<\alpha<1$ occurs
in systems such as transport in biological cells at the micron- and submicron
scales \cite{etoc18,scot23,weis04,gold06,wong04}, for water in the brain of
mice or bovine nasal cartilage \cite{lian16,yu23}, or chemical tracer diffusion
in porous media \cite{yan22,yan23,brian}. Conversely, superdiffusion with
$\alpha>1$ has been observed, i.a., for diffusion in the interstellar medium on
galactic scales \cite{litv17}, animal movement \cite{meyer23,hump10}, endogenous
cellular vesicles \cite{reve17}, and solute mixing in groundwater \cite{lian23a,
lian20}.

Anomalous diffusion can be effectively captured by various stochastic
models tailored to the specific system \cite{metz14,mura08,schu14,wang22}.
Among these, the continuous time random walk (CTRW) \cite{mont65} and
fractional Brownian motion (FBM) \cite{kolm40,mand68} are two prominent
processes commonly used to describe anomalous diffusion across a broad range
of systems. CTRWs are generalized random walks, in which the motion is defined
in terms of a waiting time PDF quantifying trapping events during the motion,
and a PDF of jump lengths \cite{mont65,scher}. FBM, in turn, is a Gaussian
process with stationary, power-law increments, exhibiting antipersistent
behavior for the Hurst exponent $H$ in the range $0<H<1/2$ and persistent
behavior for $1/2<H<1$. FBM in the limit $H=1/2$ corresponds to standard BM.
FBM is frequently used to model subdiffusion in viscoelastic environments such
as cellular cytoplasm and complex liquids \cite{sabr20,jeon11,jeon12}, as well as
superdiffusion in, e.g., amoeboid cells \cite{krap19}. FBM is also used to model
the density profile of nerve fibers in mouse brains \cite{janu20}.

A number of SPT datasets \cite {sabr20,pier18,ayaz21,bust22,wagn23} demonstrate
that the statistical properties of the motions originate from multiple underlying
mechanisms. Therefore, it is essential to employ mixed processes that exhibit
diverse statistical characteristics. Recently, hybrid stochastic processes in
the form of generalizations of FBM, have been proposed. Examples include FBM
combined with CTRW characteristic \cite{fox21,lian23}, FBM combined with
heterogeneous diffusion processes \cite{wang20}, FBM with nonlinear
transformations in time or space utilized to describe anomalous diffusion with
varying scaling behaviors \cite{metz23}, as well as FBM with stochastic
diffusivity \cite{wang20b,adri24,sabr20}. Additionally, random anomalous
exponents were explored in particle ensembles of FBM \cite{han20,balc22}.

An alternative extension of FBM can be achieved through subordination. Recent
research in geophysics has proposed a model in which hydraulic conductivity
is described by subordinating FBM to a gamma process, known as fractional
Laplace motion (FLM) \cite{kozu06,meer04,gant09,gajd17}. Introduced by
Meerschaert and colleagues \cite{meer04}, FLM combines FBM as the parent process
with a gamma process transforming the operational time of the parent process to
a new laboratory time. The gamma process is a strictly increasing L{\'e}vy process
\cite{noor09,lawl04,mada98} whose increments follow a gamma distribution. The gamma
process was first applied by Moran to model water flow into a dam \cite{mitc78},
and it is now widely used in different fields, such as maintenance \cite{noor09},
finance \cite{mada98}, or degradation \cite{lawl04}. In the context of subordination
theory \cite{boch62}, it represents the trading time or volume in financial
applications and the number of depositional features encountered over a distance
in hydrology \cite{kozu06}. FLM has been applied successfully in hydraulic
conductivity \cite{meer04,molz07}, subsurface hydrology \cite{molz06} and sediment
transport \cite{gant09}. Unlike classic FBM or most generalized versions,
where the PDFs remain Gaussian or non-Gaussian across all time scales,
however, in FLM, the PDFs vary with time: at short times the PDF is non-Gaussian,
whereas as the time scale increases, it gradually approaches a Gaussian PDF
as quantified by the kurtosis \cite{kozu06}.

We here go one step further and study FLM in the presence of a constant
drift. The drift usually arises due to constant force acting on a particle
and results in ballistic motion \cite{kuba21,kras22,comp97,metz98}. Drift
effects are observed in the transient current in an amorphous material
with an electric field \cite{rude78}, drying of porous media with temperature
and pressure gradients \cite{xiao08}, fluctuating interactions in gel
electrophoresis \cite{yarm97}, or hydrodynamic dispersion in a flow field
\cite{redn87}. In this paper, we consider two distinct drift mechanisms:
external drift, which influences the subordinated, composite process;
and internal drift, which only impacts the parent process. We analyze the
statistical properties of the process under two types of drift and explore
the complex structure of the PDF as well. We find that over time, a region
emerges, in which the PDF follows a Gaussian distribution, while outside this
region, it follows a non-Gaussian distribution. Still, due to the small amplitude
of the tail regions, the kurtosis assumes the value of a Gaussian PDF.

This paper is organized as follows. In Sec.~\ref{sec2}, we recall the
characteristic properties of FLM. Secs.~\ref{sec3} and \ref{sec4} are
dedicated to deriving the statistical properties of FLM with external and
internal drift respectively---including the moments, MSD, and the kurtosis. We
also analyze the intricate structure of the PDFs and the non-Gaussian
behaviors. Sec. \ref{sec5} provides the main conclusions.

\section{Statistical properties of FLM} 
\label{sec2}

FLM $x(t)\equiv B_H(s(t))$ is a process resulting from the subordination of FBM
to a gamma process,
\begin{equation}
\frac{dx(s)}{ds}=\zeta_H(s),\quad
\frac{ds(t)}{dt}=\varepsilon(s),
\label{eq-flm}
\end{equation}
in the subordination notation in terms of coupled Langevin
equations introduced in \cite{foge94, eule}. Here the parent process $B_H(s)$ is conventional FBM running on the operational
time $s$ and driven by the fractional Gaussian noise $\zeta_H(s)$ \cite{mand68,
mishura} with zero mean and autocovariance function (ACVF) 
\begin{equation}
\label{fgn}
\langle\zeta_H(s_1)\zeta_H(s_2)\rangle\sim 2K_H H(2H-1)|s_1-s_2|^{2 H-2}, 
\end{equation}
where $K_H$ is the generalized diffusion coefficient with physical dimension
$[K_H]=\mathrm{length}^2/\mathrm{time}^{2H}$. The PDF of the FBM parent process
$B_H(s)$ is Gaussian,
\begin{equation}
\label{eq-pdf-fbm}
G(x,s)=\frac{1}{\sqrt{4\pi K_{H} s^{2H}}}\exp\left(-\frac{x^2}{4K_Hs^{2H}}\right),
\end{equation}
and the associated MSD is given by
\begin{equation}
\langle x^2(s)\rangle=2K_Hs^{2H}.    
\end{equation}

The subordinator $s(t)$ is the L\'evy-Gamma process that is the process with independent stationary  increments \cite{chen19}. The PDF of the subordinator is thus given by
\begin{equation}
\label{eq-pdf-gamma}
h(s,t)=\frac{\lambda^{\gamma t}}{\Gamma(\gamma t)}s^{\gamma t-1}e^{-\lambda s},
\end{equation}
where $\lambda$ and $\gamma$ are both parameters of physical dimension $\mathrm{
time}^{-1}$. Finally, $\varepsilon(t)$ is the associated gamma noise.
Without loss of generality, we use the variable transform $\hat{x}=x/
(2K_H\gamma^{2H})^{1/2}$, $\hat{s}=\lambda s$, $\hat{t}=\gamma t$ in
Eqs.~(\ref{eq-flm}), (\ref{eq-pdf-fbm}) and (\ref{eq-pdf-gamma}), to arrive at
the corresponding dimensionless forms. In the following we drop the hat notation
for simplicity.

The PDF of the subordinated (or composite) process FLM $x(t)=B_H(s(t))$ is then
given in terms of the subordination integral \cite{feller,chec21,foge94,bark01,
main12}
\begin{equation}
\label{eq-pdf-flm}
P(x,t)=\int_0^{\infty}G(x,s)h(s,t)ds, 
\end{equation}
where $G(x,s)$ and $h(s,t)$ denote the dimensionless PDFs of the parent process
$B_H(s)$,
\begin{equation}
G(x,s)=\frac{1}{\sqrt{2\pi s^{2H}}}\exp\left(-\frac{x^2}{2s^{2H}}\right)
\end{equation}
and of the subordinator $s(t)$
\begin{equation}
h(s,t)=\frac{1}{\Gamma(t)}s^{t-1}e^{-s}.
\end{equation}

Using the integral representation of the Fox $H$-function (Eq.~(1.53) in
Ref.~\cite{mathai}) with parameters $a,b,c>0$,
\begin{eqnarray}
&&\int_0^{\infty}t^{a-1}\exp\left(-bt-ct^{-\rho}\right)dt\nonumber\\
\hspace*{-4cm}&=&\frac{1}{\rho b^a} H_{0,2}^{2,0}\left[ bc^{1/\rho}\left|
\begin{array}{l}\rule{1.2cm}{0.02cm}\\(a,1),(0,1/\rho)\end{array}\right.\right],
\end{eqnarray}
the PDF \eqref{eq-pdf-flm} of FLM can be expressed via the $H$-function as
\begin{eqnarray}
\nonumber
P(x,t)&=&\int_0^{\infty}\frac{s^{t-H-1}}{\sqrt{2\pi} \Gamma(t)}\exp\left(-\frac{
x^2}{2s^{2H}}-s\right)ds\\
\nonumber
&&\hspace*{-1.6cm}=\frac{1}{2H\sqrt{2\pi}\Gamma(t)}\\
&&\hspace*{-1.6cm}\times H_{0,2}^{2,0}\left[\left(\frac{
|x|}{\sqrt{2}}\right)^{1/H}\left|\begin{array}{l}\rule{1.2cm}{0.02cm}\\(t-H,1),
(0,1/(2H))\end{array}\right.\right].
\label{eq-pdf-flm-new}
\end{eqnarray} 

In particular, when $H=1/2$, using the relation between the $H$-function and the
Bessel function (Eq.~(1.128) in Ref.~\cite{mathai}), the PDF (\ref{eq-pdf-flm-new})
of FLM becomes
\begin{equation}
P(x,t)=\frac{\sqrt{2/\pi} }{\Gamma(t)}\left(\frac{|x|}{\sqrt{2}}\right)^{t-1/2}
K_{t-1/2}(\sqrt{2}|x|),
\label{bessel}
\end{equation}
where $K_\alpha(\cdot)$ is the modified Bessel function of the third kind with
index $\alpha$. In the limit $t=1$ we recover the standard Laplace distribution
$P(x,t=1)=2^{-1/2}\exp(-\sqrt{2}|x|)$. This key property bestows the process its
name, Laplace motion. In the long time limit $t\gg1$, following the asymptotic
behavior of the modified Bessel function for large order $\mu$, given in
App.~\ref{app-a} with $z>0$, we find the approximations
\begin{equation}
\label{bessel-approx}
K_\mu(z)\sim\left\{\begin{array}{ll}\displaystyle\frac{\Gamma(\mu)}{2}\left(
\frac{z}{2}\right)^{-\mu}\exp\left(-\frac{z^2}{4\mu}\right),&z\ll\mu\\[0.4cm]
\displaystyle\sqrt{\frac{\pi}{2z}}e^{-z},&z\gg\mu\end{array}\right.,
\end{equation}
we find that the PDF follows a Gaussian form
\begin{eqnarray}
P(x,t)\sim\frac{1}{\sqrt{2\pi t}}\exp\left(-\frac{x^2}{2t}\right)
\end{eqnarray}
in the regime $|x|\ll t$, and the non-Gaussian form
\begin{eqnarray}
P(x,t)\sim\frac{2^{-t/2}|x|^{t-1}}{\Gamma(t)}\exp\left(-\sqrt{2}|x|\right),
\label{eq-pdf-flm-exponetial-h=1/2}
\end{eqnarray}
in the regime $|x|\gg t$.

For arbitrary $H$, the PDF Eq.~(\ref{eq-pdf-flm-new}) of FLM also has a distinct
non-Gaussian behavior. Namely, for large $|x|$, the PDF (\ref{eq-pdf-flm-new})
asymptotically reads \cite{kozu06}
\begin{equation}
\label{eq-pdf-flm-exponetial}
P(x,t)\sim a|x|^{(2t)/(1+2H)-1}\exp\left(-b|x|^{\frac{2}{1+2H}}\right),
\quad|x|\to\infty,
\end{equation}
where the factors $a=[\sqrt{1+2H}\Gamma(t)]^{-1}H^{t/(1+2H)-1/2}$ and $b=(1+2H)/2
H^{-2H/(1+2H)}$. This non-Gaussian PDF (\ref{eq-pdf-flm-exponetial}) may also be
obtained via applying the approximation of the $H$-function for large |x| as given
in Eq.~(1.108) in \cite{mathai}. In particular, this result is consistent with the
special case $H=1/2$ in Eq.~(\ref{eq-pdf-flm-exponetial-h=1/2}) .

At long times $t$, the PDF Eq.~(\ref{eq-pdf-flm-new}) can be approximated by the
Gaussian PDF 
\begin{equation}
\label{eq-pdf-flm-gaussian}
P(x,t)\sim\frac{1}{\sqrt{2\pi t^{2H}}}\exp\left(-\frac{x^2}{2t^{2H}}\right),\quad
t\to\infty.
\end{equation}
The derivation of this asymptotic behavior for the PDF, using the properties of
the $H$-function \cite{prud}, is provided in App.~\ref{app-b}.

The moments of FLM in subordination form are given by 
\begin{equation}
\label{moment}
\langle x^n(t)\rangle=\int_0^{\infty}\langle x^n(s)\rangle h(s,t)ds.
\end{equation}
Thus one may immediately obtain the MSD (identical to the second moment) as
\begin{equation}
\label{msd}
\langle x^2(t)\rangle=\frac{\Gamma(2H+t)}{\Gamma(t)}.
\end{equation}
In particular, when $H=1/2$, FLM displays the linear growth $\langle x^2(t)
\rangle=t$ of the MSD at all times. 
Using the approximation of gamma function (Eq.~(6.1.47) in \cite{abra72}),
\begin{equation}
\label{gamma-appro}
z^{b-a}\frac{\Gamma(z+a)}{\Gamma(z+b)}\sim1+\frac{(a-b)(a+b-1)}{2z}+o(z^{-1}),
z\to\infty,
\end{equation}
we obtain the MSD of FLM at long times,
\begin{equation}
\label{msd-appro}
\langle x^2(t)\rangle\sim t^{2H}.   
\end{equation}

With the fourth-order moment
\begin{equation}
\langle x^4(t)\rangle=3\frac{\Gamma(4H+t)}{\Gamma(t)},
\end{equation}
the kurtosis of FLM $\kappa$ reads \cite{kozu06}
\begin{equation}
\kappa=\frac{\langle x^4(t)\rangle}{\langle x^2(t)\rangle^2}=3\frac{\Gamma(4H
+t)\Gamma(t)}{\Gamma^2(2H+t)}
\label{eq-kurtosis-flm}.
\end{equation}
At short times, using the approximation $\Gamma(t)\sim1/t$, the kurtosis has
the limiting form
\begin{equation}
\label{kur-appro-st}
\kappa\sim\frac{3\Gamma(4H)}{\Gamma(2H)}\times\frac{1}{t},
\end{equation}
implying that in this limit $t\to 0$, the kurtosis assumes an infinite value.
At long times, considering terms up to $1/t$ in the leading order expansion of
the gamma function in Eq.~(\ref{gamma-appro}), the approximate kurtosis yields 
in the form
\begin{eqnarray}
\label{kur-appro} 
\kappa\sim3\times\left(1+\frac{4H^2}{t}\right).
\end{eqnarray}
Thus, the kurtosis approaches the Gaussian limit $\kappa=3$ with as inverse
power of time. For $H<1/2$, the kurtosis decays more rapidly than for $H>1/2$.

\begin{figure}
\includegraphics[width=0.8\linewidth]{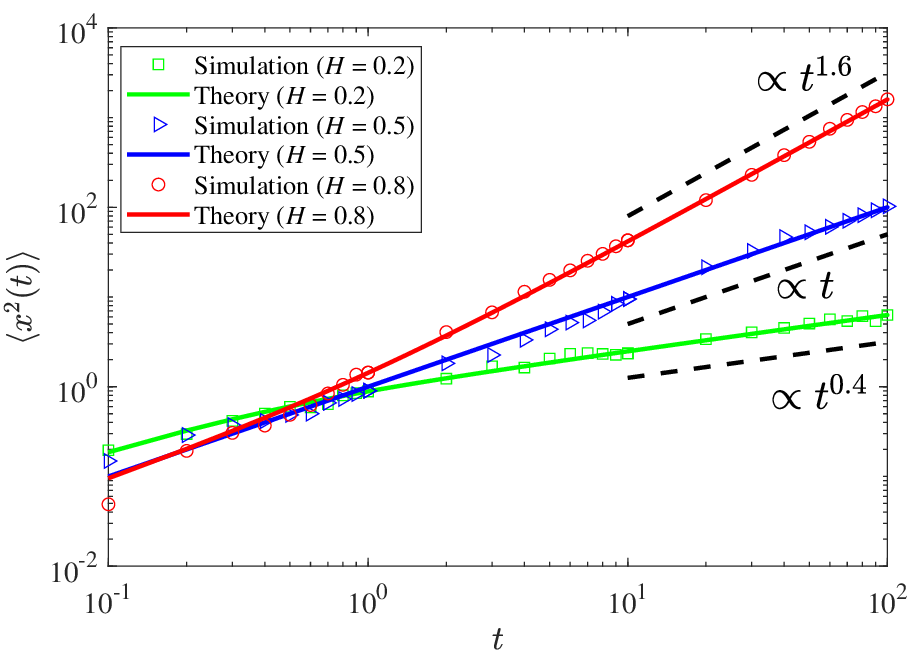}
\caption{Simulations (circles, inverted triangles, rectangles) and theoretical
results (solid curves) from Eq.~\eqref{msd} of the MSD (\ref{eq-flm}) for FLM
for subdiffusive ($H=0.2$), normal-diffusive ($H=0.5$) and superdiffusive ($H
=0.8$) cases. Parameters: the trajectory length is $T=10^2$, the simulation
time step is $dt=0.1$, and the number of trajectories is $N=300$. These
parameters are consistently kept across all figures, unless stated otherwise.}
\label{fg-msd}
\end{figure}

\begin{figure}
\includegraphics[width=0.8\linewidth]{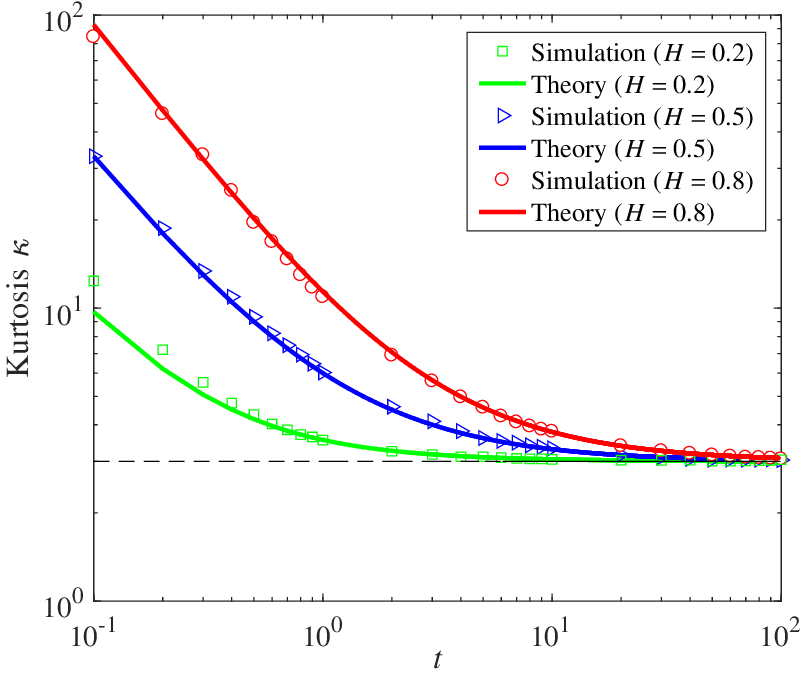}
\caption{Kurtosis (\ref{eq-flm}) for FLM. The theoretical result
\eqref{eq-kurtosis-flm} and the long time approximation
(\ref{kur-appro}) are in nice agreement with the simulations results.
The dashed line represents the Gaussian limit $\kappa=3$.}
\label{fg-kurtosis} \end{figure}

The MSD and the kurtosis of FLM from stochastic simulations for different values
of $H$ are shown in Figs.~\ref{fg-msd} and \ref{fg-kurtosis}. In both cases
the analytical results are in nice agreement with the simulations. The MSD
exhibits the power-law scaling $t^{2H}$ at long times. The kurtosis diverges
as time approaches zero, and converges to $\kappa=3$ in the long time limit,
indicating the Gaussian nature of the FLM process at long times. Note also
the discussion of the peculiar shape of the PDF below.

The increments of the FLM process are given by
\begin{equation}
x_\Delta(t)=x(t+\Delta)-x(t)=B_H(s(t+\Delta))-B_H(s(t)).
\end{equation}
On their basis and using the ACVF \cite{kozu06}, the mean-squared increment (MSI)
of FLM is given by
\begin{eqnarray}
\langle x^2_\Delta(t)\rangle=\frac{\Gamma(2H+\Delta)}{\Gamma(\Delta)}.
\end{eqnarray}
The MSI is stationary and solely depends on the lag time. Moreover, it is identical
to the MSD (\ref{msd}). At large $\Delta$ the MSI reads
\begin{equation}
\label{msi-appro}
\langle x^2_\Delta(t)\rangle\sim\Delta^{2H}.  
\end{equation}

The ACVF of the two increments $x(t_1+\Delta)-x(t_1)$ and $x(t_2+\Delta)-x(t_1)$
follows, for any $t_1,t_2>0$, in the form
\begin{eqnarray}
\nonumber
\langle x_\Delta(t_1)x_\Delta(t_2)\rangle&=&\frac{1}{2}\left(\frac{\Gamma(2H+|t_2
-t_1+\Delta|)}{\Gamma(|t_2-t_1+\Delta|)}\right.\\
&&\left.\hspace*{-3.2cm}+\frac{\Gamma(2H+|t_1-t_2+\Delta|)}{\Gamma(|t_1-t_2+\Delta|)}
-2\frac{\Gamma(2H+|t_1-t_2|)}{\Gamma(|t_1-t_2|)}\right).
\end{eqnarray} 
Moreover, when $|t_2-t_1|\gg1$, the ACVF has the asymptotic power-law form
\begin{eqnarray}
\label{acf-appro}
\langle x_\Delta(t_1)x_\Delta(t_2)\rangle\sim H(2H-1)\Delta^2|t_2-t_1|^{2H-2}.
\end{eqnarray}
This result is similar to the fractional Gaussian noise ACVF (\ref{fgn}).

One can also define the fractional Laplace noise
\begin{equation}
\label{FLN}
\eta(t)=\frac{x_\Delta(t)}{\Delta}.
\end{equation}
Then the ACVF of $\eta(t)$ with $|t_2-t_1|\gg1$ reads
\begin{equation}
\langle\eta(t_1)\eta(t_2)\rangle\sim H(2H-1)|t_2-t_1|^{2H-2}.
\end{equation}
When $t_1=t_2=t$, the variance of the fractional Laplace noise is
\begin{equation}
\langle\eta^2(t)\rangle=\frac{\Gamma(2H+\Delta)}{\Gamma(\Delta)\Delta^2}. 
\end{equation}

According to the statistical properties of the MSD (\ref{msd-appro}), the MSI
(\ref{msi-appro}), the ACVF of the increments (\ref{acf-appro}), and the kurtosis
(\ref{kur-appro}), FLM is identical to FBM in the long time limit.

\section{External drift acting on the composite FLM process}
\label{sec3}

To assess the influence of an external drift on the FLM dynamics, we first
study the case when the drift acts on the composite FLM process $x(t)=B_H(s(
t))$. The associated Langevin equation is then given by
\begin{equation}
\label{FLM-drift_2}
\frac{dx(t)}{dt}=\eta(t)+v,
\end{equation} 
where the drift $v$ is a constant and $\eta(t)$ is the fractional Laplace noise
defined in Eq.~(\ref{FLN}), obeying $\eta(t)dt=\left[B_H(s(t+dt))-B_H(s(t))
\right]$. 

\begin{figure}
(a)\includegraphics[width=0.8\linewidth]{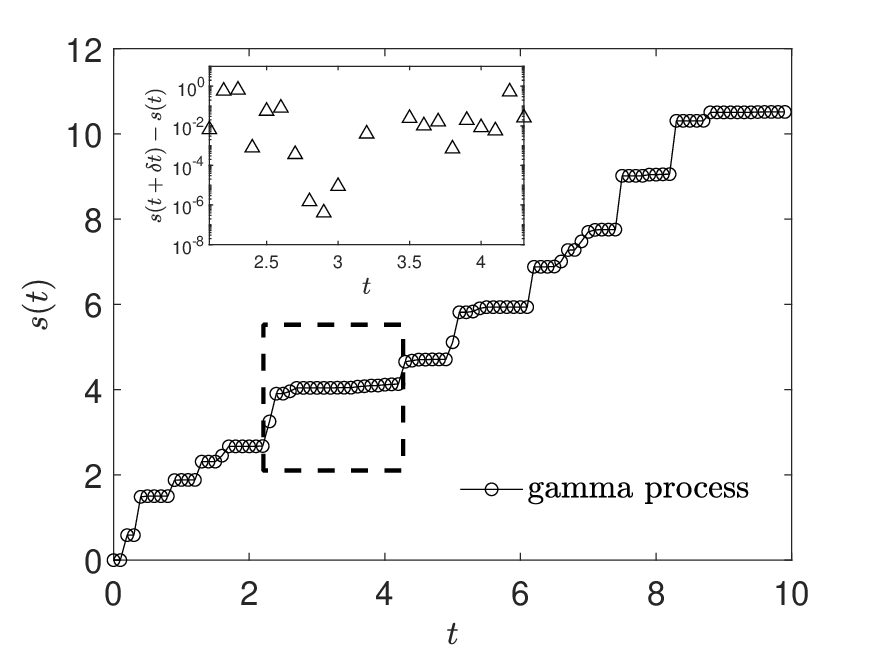}
(b)\includegraphics[width=0.8\linewidth]{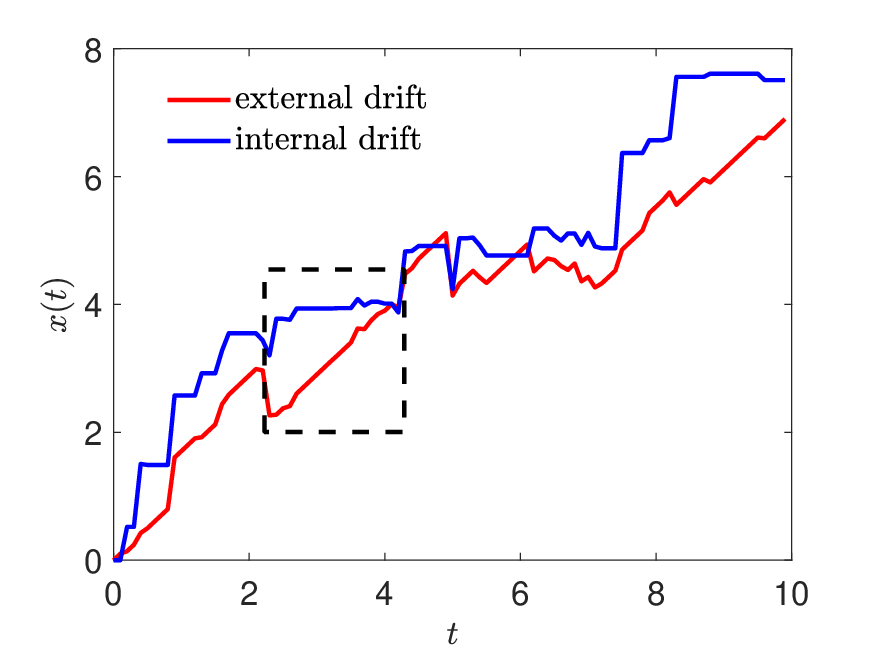}
\caption{Trajectories of (a) the subordinator gamma process $s(t)$ and (b) the
composite FLM process $x(t)$ of duration $T=10$ in the presence of an external
drift (red curve, Eq.~\eqref{FLM-drift_2}) and an internal drift (blue curve,
Eq.~\eqref{FLM-drift}). In panel (a), each time step of the corresponding
trajectory of the gamma process $s(t)$ is represented by circle. During a
specific time interval $[t,t+\Delta]$, marked by the dashed rectangle, the
gamma process appears to be trapped but is actually displaying variations in
its trajectory. The corresponding small but finite increments of $s(t)$ are
depicted as the triangles in the inset within the same time interval.
During the same time period, also marked by a dashed rectangle in panel (b),
the trajectory of FLM with internal and external drift has distinct behaviors:
for internal drift the trajectory exhibits small increments, while the for the
external drift case, the motion shows a distinct slope due to the drift.
Parameters: $H=1/2$, $dt=0.1$.}
\label{trajectory}
\end{figure}

Figure \ref{trajectory} illustrates the distinct trajectories of FLM with
either external or internal drift (discussed in the
next Section) $v=1$. In panel (a), during a time interval $[t,t+\Delta]$, in
which the gamma process $s(t)$ progresses by the small increment $s_\Delta(t)
=s(t+\Delta)-s(t)\ll1$ (as evidenced by the blowup corresponding to the
highlighted dashed rectangle). As seen in panel (b), FLM with internal drift,
acting exclusively on the parent process (\ref{FLM-drift}), experiences a small
change over the same period as the increment of the process $x(t)$ during the
same time window, given by $x_\Delta(t)=B_H(s+s_\Delta)-B_H(s+s_\Delta)=(v+
\zeta_H(s))\times s_\Delta\ll1$. In contrast, FLM with external drift,
Eq.~(\ref{FLM-drift_2}), shows a distinct slope, as the drift acts on the
composite process. The latter results in an increment $x_\Delta(t)=x(t+\Delta)
-x(t)\approx B_H(s(t+\Delta))-B_H(s(t))+v\Delta$.

Analogous to the PDF \eqref{eq-pdf-flm-new} of free FLM, the PDF of FLM with
external drift can be obtained as
\begin{eqnarray}
\nonumber
P(x,t)&=&\int_0^{\infty}\frac{s^{t-H-1}}{\sqrt{2\pi}\Gamma(t)}\exp\left(-\frac{
(x-vt)^2}{2s^{2H}}-s\right)ds\\
\nonumber
&&\hspace*{-1.8cm}=\frac{1}{2H\sqrt{2\pi}\Gamma(t)}\\
&&\hspace*{-1.8cm}\times H_{0,2}^{2,0}\left[\left(\frac{|x-vt|}{\sqrt{2}}\right)
^{1/H}\left|\begin{array}{l}\rule{1.2cm}{0.02cm}\\(t-H,1),(0,1/(2H))\end{array}
\right.\right].
\label{pdf-flm-drift_2}
\end{eqnarray}
When $H=1/2$, the PDF \eqref{pdf-flm-drift_2} can be expressed in terms of the
modified Bessel function in the form
\begin{equation}
\label{pdf-flm-drift-exact1}
P(x,t)=\frac{\sqrt{2/\pi}}{\Gamma(t)}\left(\frac{|x-vt|}{\sqrt{2}}\right)^{t-1/2}
K_{t-1/2}(\sqrt{2}|x-vt|).
\end{equation}
We note that for all $H$ values the PDF of FLM with external drift is shifted
by $vt$ as compared to the PDF of the free FLM in Eq.~(\ref{eq-pdf-flm-new}).
Consequently, the PDF remains symmetric and is
centered at $x=vt$.

Due to this Galilei variant behavior the first moment yields in the form
\begin{equation}
\label{mu1}
\langle x(t)\rangle=vt.
\end{equation}
The second moment
\begin{equation}
\label{second_drift_2}
\langle x^2(t)\rangle=\frac{\Gamma(2H+t)}{\Gamma(t)}+v^2t^2
\end{equation}
involves the ballistic shift $v^2t^2$, and thus the MSD (the second central
moment) has the same form as the original FLM process,
\begin{equation}
\langle\Delta x^2(t)\rangle=\frac{\Gamma(2H+t)}{\Gamma(t)}.
\label{msd_3}
\end{equation}
We note that at long times, the second moment scales as
\begin{equation}
\label{msd_drift_2_appro}
\langle x^2(t)\rangle\sim t^{2H}+v^2t^2,
\end{equation}
and in this limit the MSD becomes
\begin{equation}
\label{drift_2_appro}  
\langle\Delta x^2(t)\rangle\sim t^{2H}.
\end{equation}

\begin{figure}
(a)\includegraphics[width=0.8\linewidth]{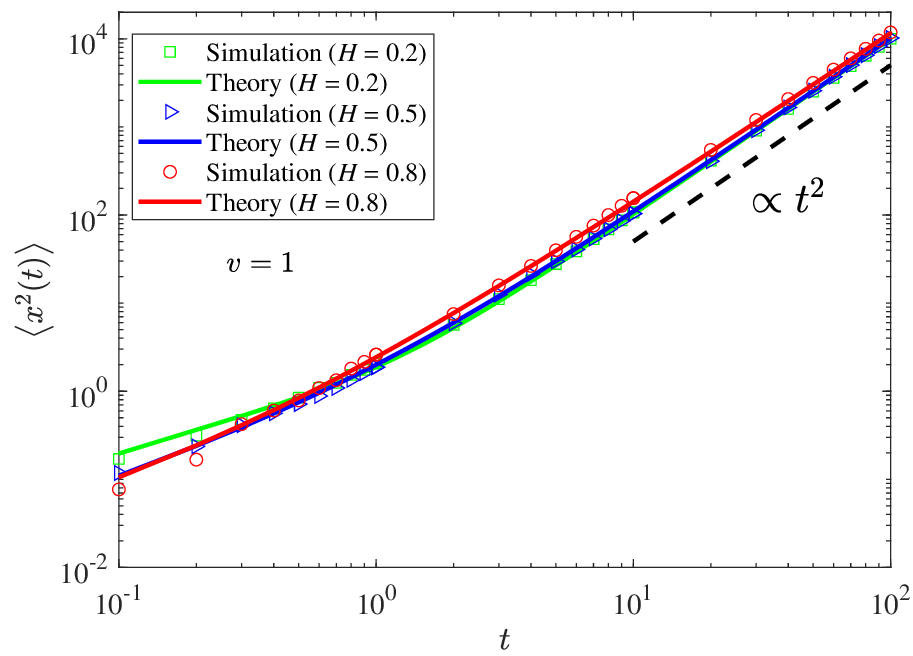}
(b)\includegraphics[width=0.8\linewidth]{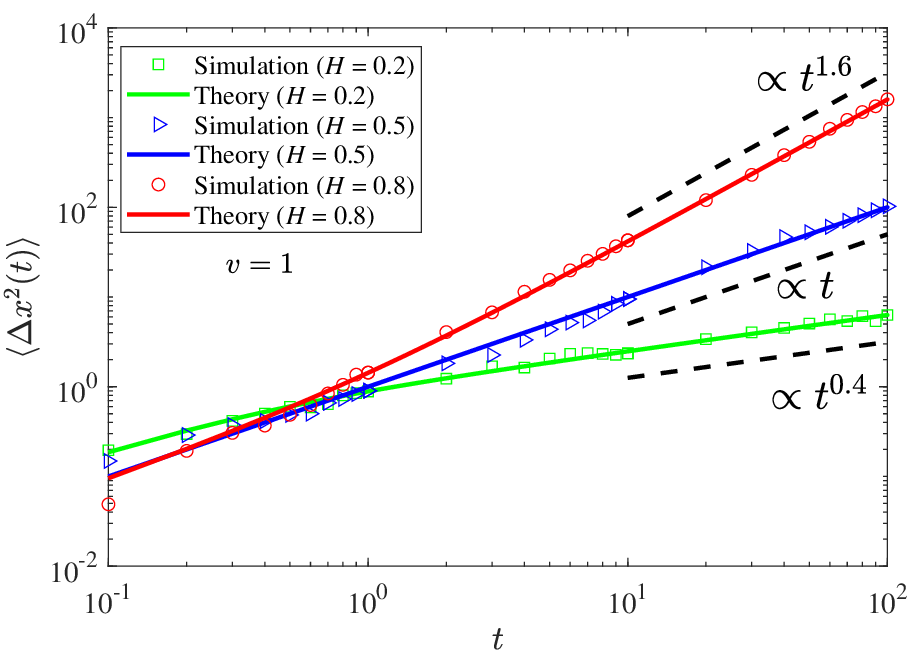}
\caption{(a) Second moment and (b) MSD for FLM with external drift $v=1$. The
theoretical results \eqref{second_drift_2} and \eqref{msd_3} nicely match the
simulations results. The external drift results in ballistic motion in the
second moment at long times and has no impact on the MSD.}
\label{fg-msd-drift_2}
\end{figure}

Figures~\ref{fg-msd-drift_2} show simulations and theoretical results for the
second moment (\ref{second_drift_2}) and the MSD (\ref{msd_3}) for FLM with
external drift $v=1$. The analytical results are in nice agreement with the
simulations. At long times, the second moment scales like ballistic motion due
to the second term in Eq.~(\ref{msd_drift_2_appro}) which dominates at long times,
while the external drift is subtracted out in the MSD (\ref{drift_2_appro}).

\begin{figure}
\includegraphics[width=0.8\linewidth]{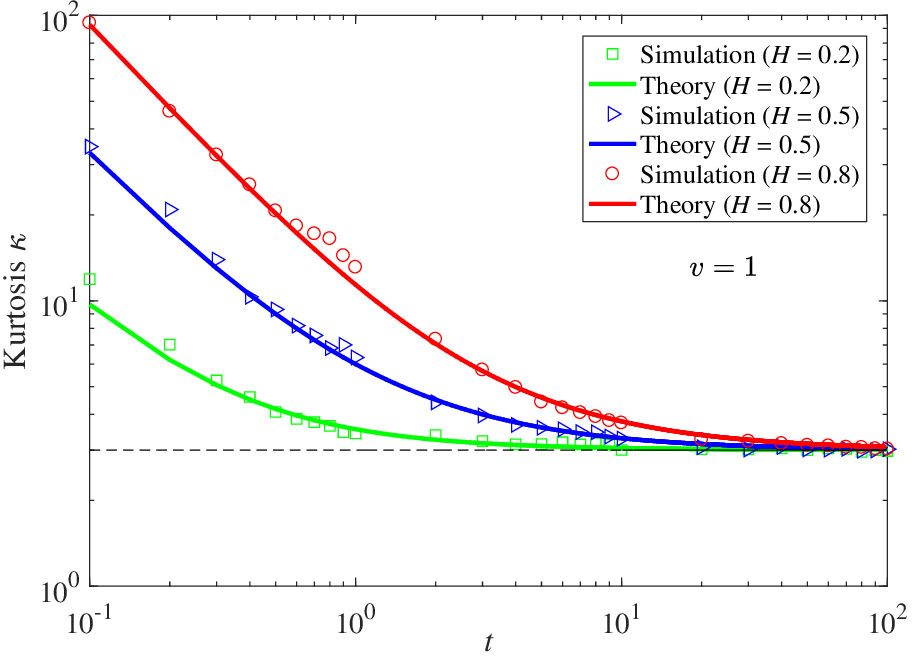}
\caption{Kurtosis for FLM with external drift $v=1$, Eq.~(\ref{FLM-drift_2}).
The dashed line represents the Gaussian value $\kappa=3$. As the external has
no impact on the kurtosis, the simulation results coincide with the theoretical
result \eqref{eq-kurtosis-flm} for the process without drift.}
\label{fg-kurtosis-drift_2} \end{figure}

As the external drift has no effect on the central moments, the kurtosis $\kappa$
of FLM with external drift has the same form as unbiased FLM, as given in
Eq.~\eqref{eq-kurtosis-flm}. Figure~\ref{fg-kurtosis-drift_2} shows simulations
results for the kurtosis of FLM with external drift, consistent with the analytical
results in Eq.~\eqref{eq-kurtosis-flm}. 

Using Stirling's formula (Eq.~(6.1.37) in \cite{abra72}),
\begin{equation}
\label{gamma_app}
\Gamma(t)\sim\sqrt{2\pi}t^{t-1/2}e^{-t},\quad t\gg1,
\end{equation}
the PDF (\ref{pdf-flm-drift_2}) can be rewritten as
\begin{equation}
\label{eq-pdf-flm-drift_2}
P(x,t)=\frac{1}{2\pi t^{t-1/2}}\int_0^{\infty}s^{-H-1}\mathrm{exp}(-\phi(s))ds,
\end{equation}
with the function 
\begin{equation}
\label{eq-fai_2}
\phi(s)=\frac{(x-vt)^2}{2s^{2H}}+s-t\ln (s)-t.
\end{equation}
Based on this approximate formulation we now discuss the asymptotic behaviors of
the PDF in the cases $v=0$ and $v\neq0$.

\subsection{The unbiased case $v=0$}

When $v=0$ the function $\phi(s)$ in Eq.~\eqref{eq-fai_2}) becomes $\phi(s)=\frac{
x^2}{2s^{2H}}+s-t\ln(s)-t$, attaining a minimum at $s=s_m$. $s_m$ may be obtained
by solving $\phi^\prime(s_m)=0$, producing
\begin{equation}
\label{eq-minima-free}
\frac{t}{s_m}+\frac{Hx^2}{s_m^{2H+1}}=1.
\end{equation}
This equation cannot be solved exactly, except for the case $H=1/2$, when the
solution reads $s_m=\frac{1}{2}(t+\sqrt{t^2+2x^2})$. However, it is still possible
to obtain an asymptotic solution under specific conditions.

In Eq.~(\ref{eq-minima-free}), it is obvious that $t/s_m<1$ and $Hx^2/s_m^{2H+1}
<1$. Moreover, if $x^2\gg t^{2H+1}$, the first term is much less than the second
term, and the minimum is attained at $s_m=\left(Hx^2\right)^{1/(2H+1)}$. Then the
standard Laplace method can be used, and one arrives at the non-Gaussian form
(\ref{eq-pdf-flm-exponetial}),
\begin{eqnarray}
\nonumber
P(x,t)&\sim&\frac{1}{2\pi t^{t-1/2}}s_m^{-H-1} \sqrt{\frac{2\pi}{\phi^{''}
(s_m)}}\exp\big(-\phi(s_m)\big)\\
&\sim&a|x|^{(2t)/(1+2H)-1}\exp\left(-b|x|^{2/(1+2H)}\right).
\label{eq-tail-free}
\end{eqnarray}
In contrast, when $x^2\ll t^{2H+1}$, the first term is much larger than the
second term in Eq.~(\ref{eq-minima-free}) and the minima is attained at $s_m=
t$. Then, the PDF has asymptotic Gaussian form
\begin{eqnarray}
\nonumber
P(x,t)&\sim&\frac{1}{2\pi t^{t-1/2}}s_m^{-H-1} \sqrt{\frac{2\pi}{\phi^{''}
(s_m)}}\exp\big(-\phi(s_m)\big)\\
&\sim&\frac{1}{\sqrt{2\pi t^{2H}}}\exp\left(-\frac{x^2}{2t^{2H}}\right).
\label{eq-gau-free}
\end{eqnarray}
This analysis demonstrates that the PDF $P(x,t)$ is Gaussian inside the interval
$(-t^{H+1/2},t^{H+1/2})$. As time progresses, this interval expands to larger
$|x|$ values. However, outside this interval, the PDF is characterized by the
non-Gaussian shape (\ref{eq-tail-free}).

\begin{figure}
(a)\includegraphics[width=0.8\linewidth]{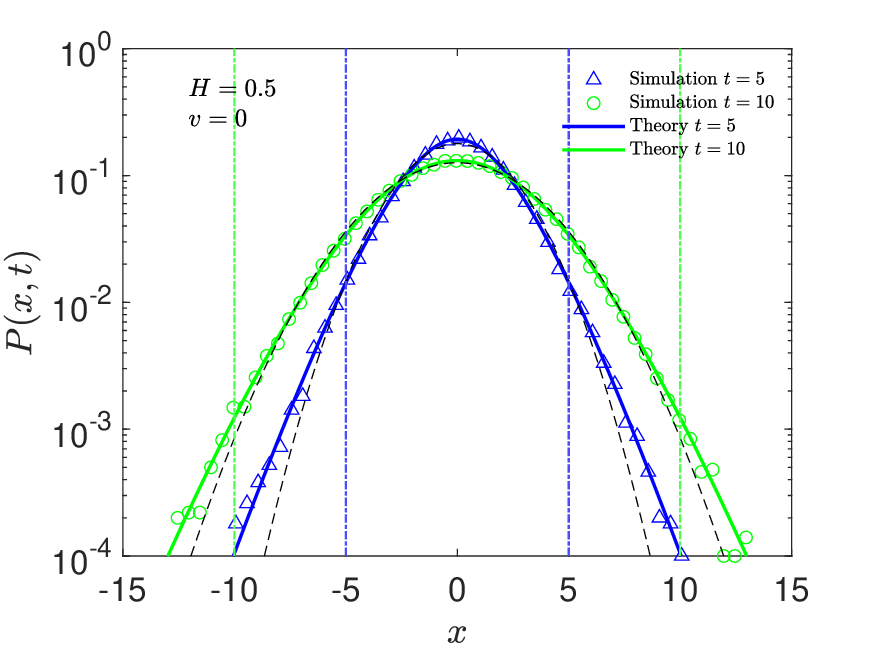}
(b)\includegraphics[width=0.8\linewidth]{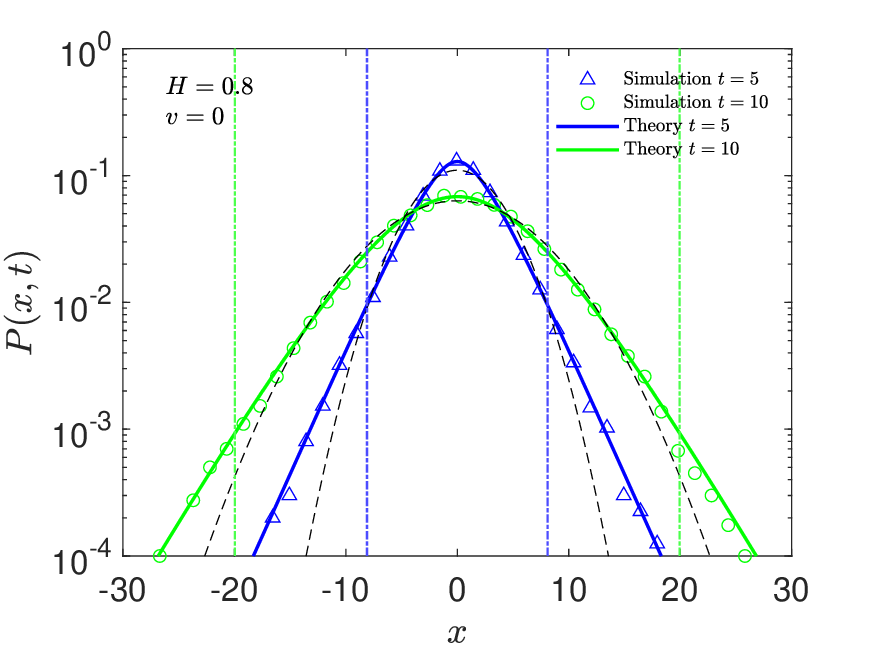}
(c)\includegraphics[width=0.8\linewidth]{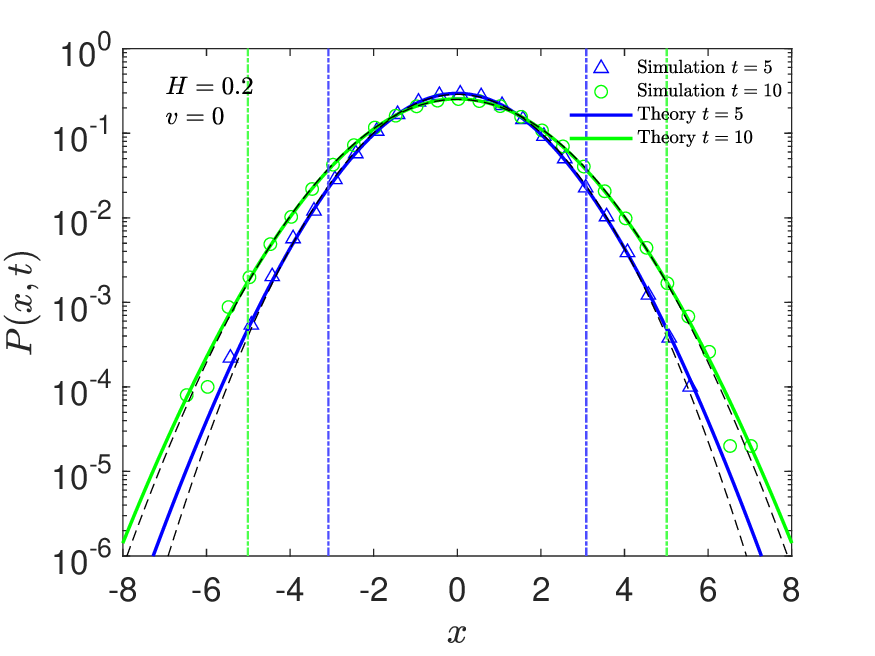}
\caption{Simulations (symbols) for the PDF of free FLM at times $t=5$, and $t=
10$ with (a) $H=0.5$, (b) $H=0.8$ and (c) $H=0.2$. The analytical PDF (colored
solid curves) correspond to Eq.~\eqref{eq-pdf-flm-new}, while the Gaussian
approximation \eqref{eq-gau-free} is represented by the dashed curves. The
colored vertical lines $x=\pm t^{H+1/2}$ are identified as boundaries separating
the Gaussian and non-Gaussian domains.}
\label{fg-pdf}
\end{figure}

Figure \ref{fg-pdf} shows the results of our analytical calculations and
stochastic simulations of the PDF for FLM without drift at three different
times for the Hurst exponents $H=0.5$, $H=0.8$ and $H=0.2$. The analytical
results agree well with the simulations in all cases. As predicted, the tails
of the PDF are increasingly suppressed by the Gaussian component as function
of time, such that the kurtosis becomes decreasingly sensitive to the extreme
non-Gaussian tails, as demonstrated in Fig.~\ref{fg-kurtosis}. 

\subsection{The biased case $v\neq0$}

When the external drift is non-zero, based on the function $\phi(s)$ from
Eq.~(\ref{eq-fai_2}) in the PDF (\ref{eq-pdf-flm-drift_2}) we can obtain a
minimum at $s=s_k$, where $s_k$ may be obtained by solving the equation
\begin{equation}
\label{eq-minima-drift_2}
\frac{t}{s_k}+\frac{H(x-vt)^2}{s_k^{2H+1}}=1.
\end{equation}

Following the same strategy as used in the case without drift, the solution is
$s_k=\frac{1}{2}\left(t+\sqrt{t^2+2(x-vt)^2}\right)$ when $H=1/2$. When $(x-vt)
^2\gg t^{2H+1}$, the first term is much less than the second term in
Eq.~(\ref{eq-minima-drift_2}), and the minimum is attained at $s_k=\left(H(x-
vt)^2\right)^{1/(2H+1)}$. Then the standard Laplace method can be used again,
and one can arrive at the non-Gaussian shape
\begin{eqnarray}
\nonumber
P(x,t)&\sim&\frac{1}{2\pi t^{t-1/2}}s_k^{-H-1}\sqrt{\frac{2\pi}{\phi^{''}(s_k)}}
\exp\big(-\phi(s_k)\big)\\
\nonumber
&\sim& a|x-vt|^{(2t)/(1+2H)-1}\\
&&\times\exp\left(-b|x-vt|^{2/(1+2H)}\right).
\label{eq-tail-drift_2}
\end{eqnarray}
In contrast, when $(x-vt)^2\ll t^{2H+1}$, the first term is much larger than
the second term in Eq.~(\ref{eq-minima-drift_2}) and the minimum is attained
at $s_k=t$. Then, the PDF has the asymptotic Gaussian form
\begin{eqnarray}
\nonumber
P(x,t)&\sim&\frac{1}{2\pi t^{t-1/2}}s_k^{-H-1} \sqrt{\frac{2\pi}{\phi^{''}
(s_k)}}\exp\big(-\phi(s_k)\big)\\
&\sim&\frac{1}{\sqrt{2\pi t^{2H}}}\exp\left(-\frac{(x-vt)^2}{2t^{2H}}\right).
\label{eq-gau-drift_2}
\end{eqnarray}

\begin{figure}
(a)\includegraphics[width=0.8\linewidth]{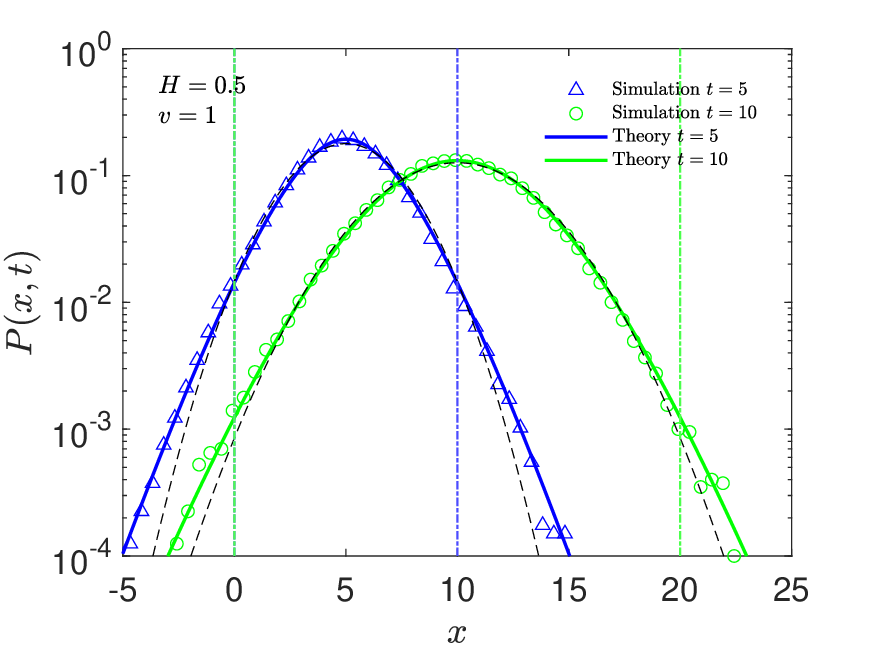}
(b)\includegraphics[width=0.8\linewidth]{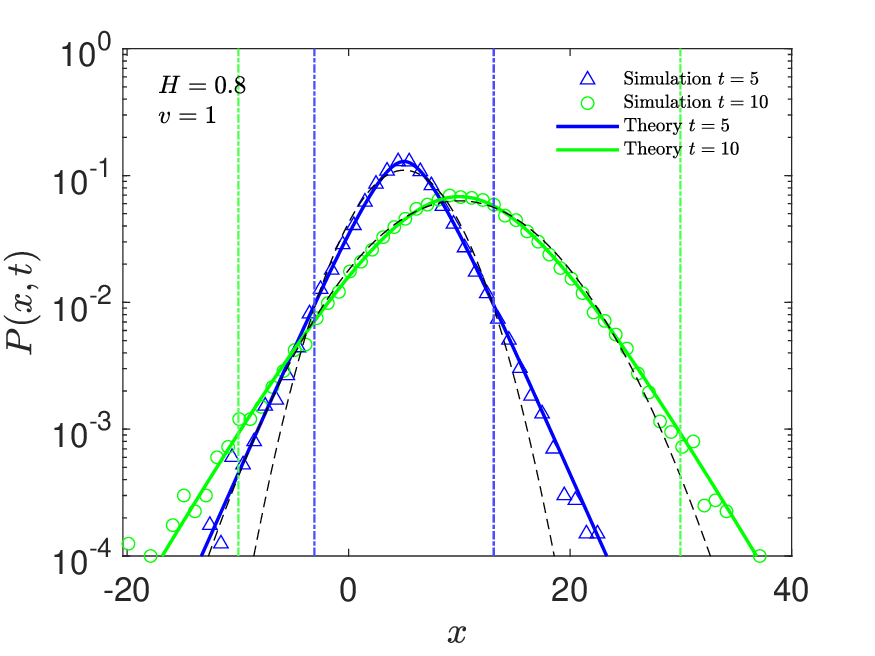}
(c)\includegraphics[width=0.8\linewidth]{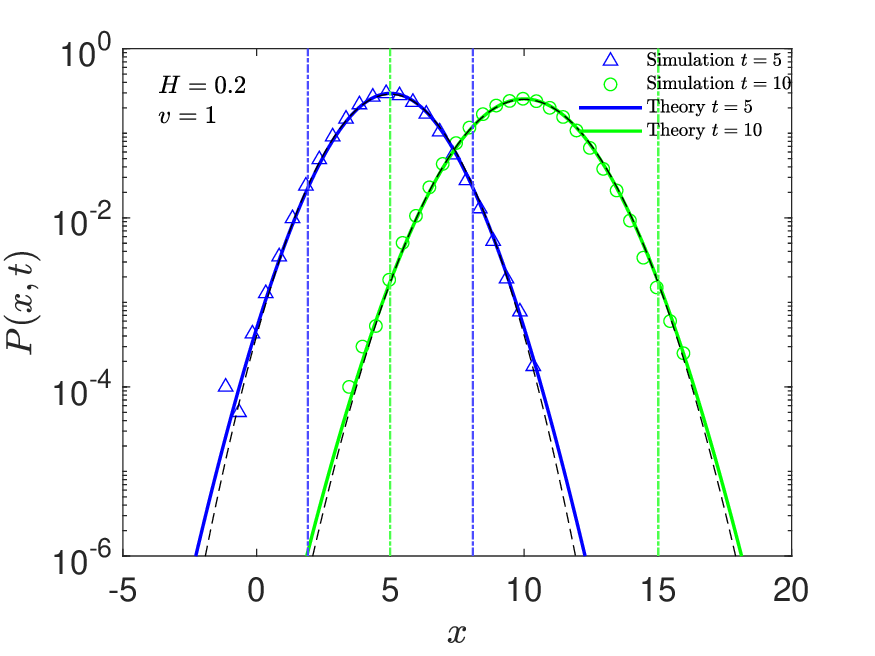}
\caption{Simulations (symbols) for the PDF of FLM with external drift $v=1$,
Eq.~(\ref{FLM-drift_2}), at $t=5$, and $t=10$ with (a) $H=0.5$, (b) $H=0.8$,
and (c) $H=0.2$. The analytical results (colored solid curves) are taken
from Eq.~\eqref{pdf-flm-drift_2}. The Gaussian approximations (dashed curves)
correspond to Eq.~\eqref{eq-gau-drift_2}. The colored vertical lines $x=vt\pm
t^{H+1/2}$ are identified as boundaries separating the Gaussian and non-Gaussian
regimes. In particular, for $H=0.5$, the left boundaries at all times $t$ are
$x=0$.}
\label{fg-pdf-drift_2}
\end{figure}

Figure \ref{fg-pdf-drift_2} shows the results of our analytical calculations
along with results from stochastic simulations of the PDF for FLM with external
drift $v=1$ acting on the subordinated process $x(t)$ for the Hurst exponents
$H=0.5$, $H=0.8$, and $H=0.2$. The analytical results agree well with the
simulations for all cases. The PDFs are symmetric and centered at $x=vt$.
Within the region $(vt-t^{H+1/2},vt+t^{H+1/2})$, one can clearly identify the
Gaussian shape from Eq.~(\ref{eq-gau-drift_2}), while the non-Gaussian behavior
(\ref{eq-tail-drift_2}) is observed outside this region.

\section{Internal drift acting solely on the parent process}
\label{sec4}

We now consider the FLM $x(t)=B_H(s(t))$ with internal constant drift $v$ solely
acting on the parent process $s$, i.e.,
\begin{equation}
\frac{dB_H(s)}{ds}=v+\zeta_H(s),\quad
\frac{ds(t)}{dt}=\varepsilon(s).
\label{FLM-drift}
\end{equation}
The integral expressions of the PDF of FLM with internal drift is then given by
\begin{equation}
P(x,t)=\int_0^{\infty}\frac{s^{t-H-1}}{\sqrt{2\pi}\Gamma(t)}\exp\left(-\frac{
(x-vs)^2}{2s^{2H}}-s\right)ds.
\label{pdf-flm-drift}
\end{equation}
In particular, for $H=1/2$, the PDF can be expressed in terms of the modified
Bessel function,
\begin{eqnarray}
\nonumber
P(x,t)&=&\frac{\sqrt{2/\pi}}{\Gamma(t)}\left(\frac{|x|}{\sqrt{v^2+2}}\right)^
{t-1/2}\\
&&\times\exp\left(vx\right)K_{t-1/2}\left(\sqrt{v^2+2}|x|\right).
\label{pdf-flm-drift-exact}
\end{eqnarray}

The first moment is again given by the linear growth in time
\begin{equation}
\label{mu}
\langle x(t)\rangle=vt,
\end{equation}
while in the second moment
\begin{eqnarray}
\label{msd_drift}
\langle x^2(t)\rangle&=&\frac{\Gamma(2H+t)}{\Gamma(t)}+v^2\frac{\Gamma(2+t)}{\Gamma(t)}\nonumber\\
&=&\frac{\Gamma(2H+t)}{\Gamma(t)}+v^2(1+t)t
\end{eqnarray}
the $v$-dependent term is given by $v^2(1+t)t$ instead of $v^2t^2$. Therefore,
a term proportional to $v^2$ remains in the second central moment 
\begin{equation}
\langle\Delta x^2(t)\rangle=\frac{\Gamma(2H+t)}{\Gamma(t)}+v^2t.
\label{msd_2}
\end{equation}
Using the approximation of the gamma function (\ref{gamma-appro}) we find the
long time limit of the second moment 
\begin{equation}
\langle x^2(t)\rangle\sim t^{2H}+v^2t^2,
\end{equation}
and the MSD
\begin{equation}
\langle\Delta x^2(t)\rangle\sim t^{2H}+v^2t.
\label{msd_22}
\end{equation}

\begin{figure}
(a)\includegraphics[width=0.8\linewidth]{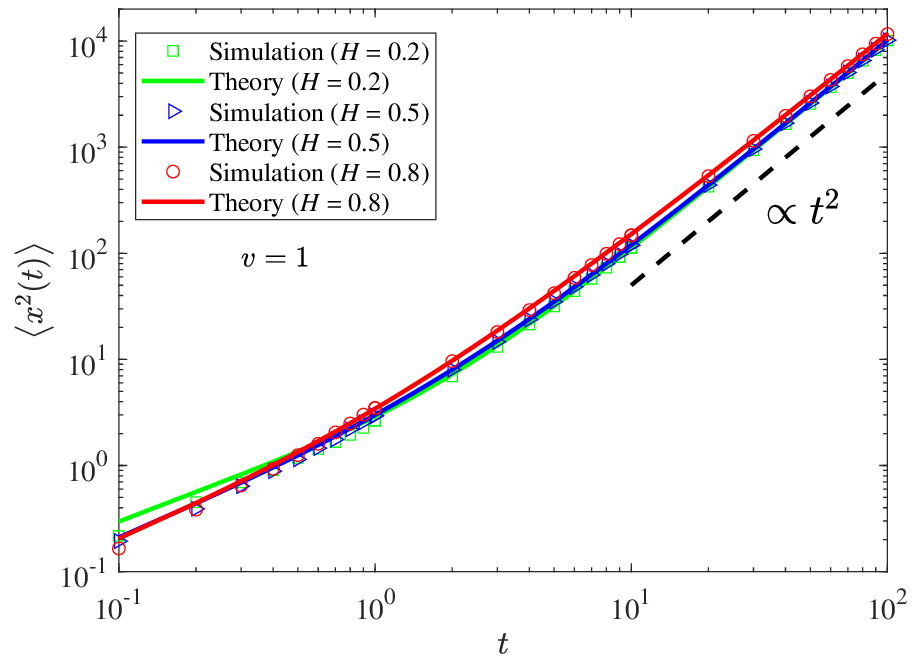}
(b)\includegraphics[width=0.8\linewidth]{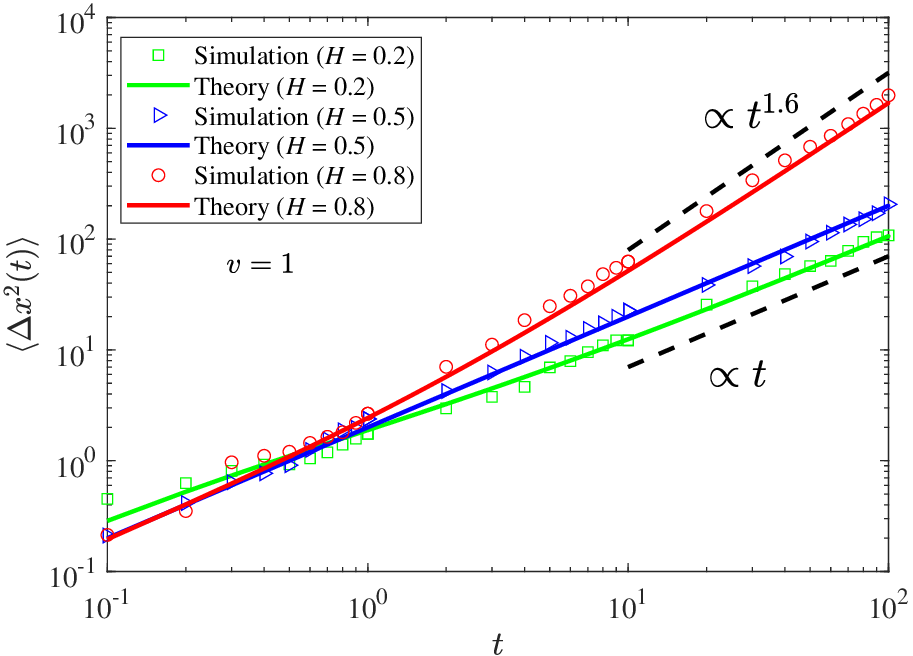}
\caption{(a) Second moment and (b) MSD for FLM with internal drift with $v=1$.
The theoretical results from Eqs.~\eqref{msd_drift} and \eqref{msd_2} agree
nicely with the results from stochastic simulations.}
\label{fg-msd-drift}
\end{figure}

Figure \ref{fg-msd-drift} displays simulations results for the second moment
for FLM with internal drift ($v=1$), Eq.~(\ref{msd_drift}), and the associated
MSD, Eq.~(\ref{msd_2}), for different values of $H$. We also show the analytical
results, which are in good agreement with the simulations. The behavior of these
moments for FLM with internal drift are distinctly different from the case of an
external drift shown in Fig.~\ref{fg-msd-drift_2}. Both types of drifts lead to
ballistic motion in the second moment at long times, as indicated by expressions
(\ref{msd_drift_2_appro}) and (\ref{msd_2}). However, while the external drift
does not affect the MSD (\ref{drift_2_appro}), the internal drift leads to an
apparent normal diffusion in the MSD (\ref{msd_22}), which is the dominant
scaling at long times for $H<1/2$.   

The kurtosis $\kappa$ can be written as the ratio
\begin{equation}
\label{eq-kurtosis-drift}
\kappa=\frac{\langle\Delta x^4(t)\rangle}{\langle\Delta x^2(t)\rangle^2}
\end{equation}
of the central moments, where $\langle\Delta x^4(t)\rangle$ is the fourth
central moment given by 
\begin{eqnarray}
\nonumber
\langle[x(t)-\langle x(t)\rangle]^4\rangle&=&3\left\{\frac{\Gamma(4H+t)}{\Gamma(
t)}+v^4(t+2)t\right.\\
&&\hspace*{-2.0cm}\left.+2v^2\frac{\Gamma(2H+t)}{\Gamma(t)}\Big[t+2H(2H+1)
\Big]\right\}.    
\end{eqnarray}
In the special case $H=1/2$ the kurtosis assumes the algebraic form
\begin{eqnarray}
\label{d_kur_h05}
\kappa=3\left(1+\frac{2}{t}-\frac{1}{(1+v^2)^2t}\right).
\end{eqnarray}
The latter approaches the Gaussian value 3 in the long time limit. The latter
result is consistent with the expression \eqref{eq-kurtosis-flm} when $v=0$. 

\begin{figure}
\includegraphics[width=0.8\linewidth]{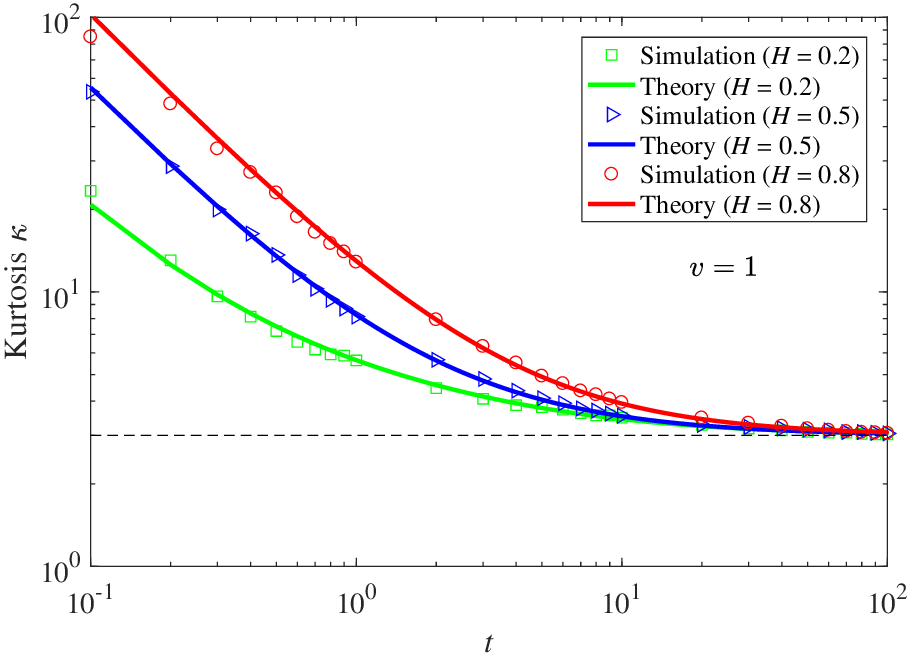}
\caption{Simulations and theoretical results, Eq.~\eqref{eq-kurtosis-drift}, for
the kurtosis for FLM with internal drift ($v=1$) for varying values of the Hurst
exponent $H$. The dashed line represents $\kappa=3$.}
\label{fg-kurtosis-drift}
\end{figure}

Figure \ref{fg-kurtosis-drift} displays the analytical results and simulations
for the kurtosis for FLM with internal drift as a function of time, based on
Eq.~\eqref{eq-kurtosis-drift}, showing good agreement. Compared to FLM without
drift shown in Fig. \ref{fg-kurtosis}, the kurtosis for FLM with internal drift
displays larger values at shorter times and approaches the Gaussian value of
$\kappa=3$ more slowly at long times, as indicated by Eq.~\eqref{d_kur_h05}.

The PDF (\ref{pdf-flm-drift}) can be rewritten in the same form we introduced
in Eq.~\eqref{eq-pdf-flm-drift_2}. Now the function $\phi(s)$ assumes the form
\begin{equation}
\label{eq-fai}
\phi(s)=\frac{(x-vs)^2}{2s^{2H}}+s-t\ln(s)-t.
\end{equation}
In what follows we discuss the asymptotic behavior of the PDF for the case
$v\neq0$.

\begin{figure}
(a)\includegraphics[width=0.8\linewidth]{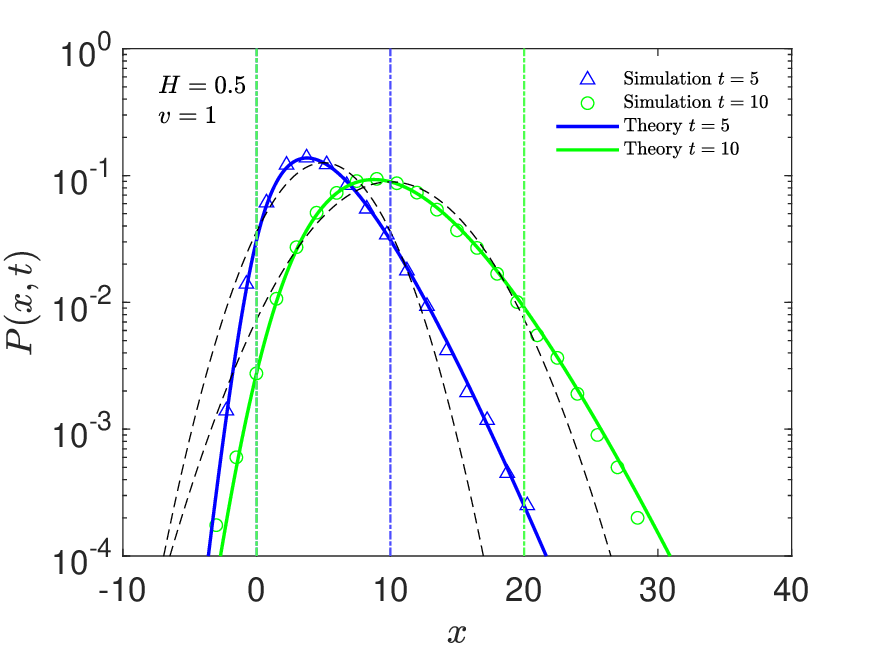}
(b)\includegraphics[width=0.8\linewidth]{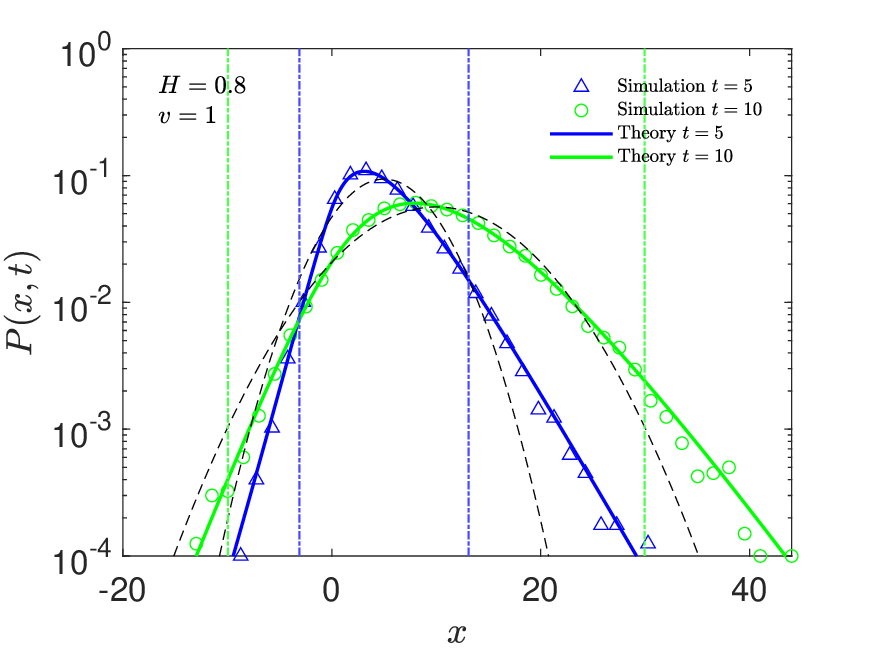}
(c)\includegraphics[width=0.8\linewidth]{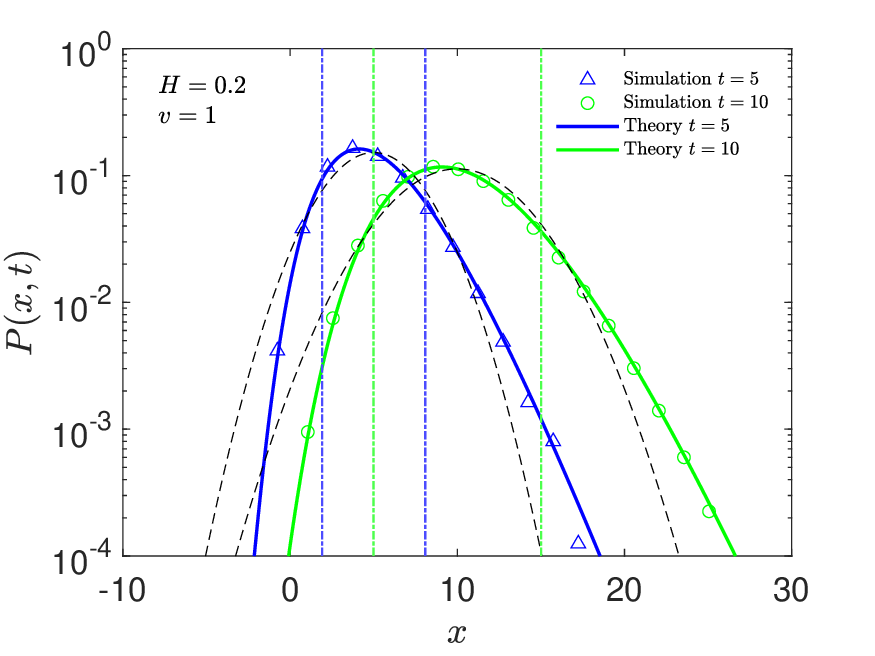}
\caption{Simulations (symbols) for the PDF (\ref{FLM-drift}) of FLM with
internal drift for $v=1$ and at times $t=5$, and $t=10$ with (a) $H=0.5$,
(b) $H=0.8$ and (c) $H=0.2$. The analytical results (colored solid curves)
correspond to Eq.~\eqref{pdf-flm-drift}. The Gaussian approximations (dashed
curves) are taken from Eq.~\eqref{eq-gau-drift}. The boundaries of the Gaussian
and non-Gaussian domains $x=vt\pm t^{H+1/2}$ are represented by the colored
vertical lines.} 
\label{fg-pdf-drift}
\end{figure}

When the drift is non-zero, the function (\ref{eq-fai}) assumes a minimum at
$s=s_n$, where $\phi^\prime(s_n)=0$. This leads is to the implicit equation
\begin{equation}
\label{eq-minima-drift}
\frac{t}{s_n}+\frac{Hx^2}{s_n^{2H+1}}+\frac{(1-2H)vx}{s_n^{2H}}+\frac{(H-1)
v^2}{s_n^{2H-1}}=1.
\end{equation}
This relation can be solved exactly when $H=1/2$, with the solution $s_n=\big(
t+\sqrt{t^2+(2+v^2)x^2}\big)/(2+v^2)$. In addition, when $x^2\gg t^2$, $s_n
\sim|x|/\sqrt{v^2+2}$. Then the standard Laplace method can be applied, and one
can estimate the PDF by the non-Gaussian distribution
\begin{eqnarray}
\nonumber
P(x,t)&\sim& \frac{1}{2\pi t^{t-1/2}}s_n^{-3/2} \sqrt{\frac{2\pi}{\phi^{''}
(s_n)}}\exp(-\phi(s_n))\\
&\sim&\overline{a}|x|^{t-1}\exp\left(-\overline{b}|x|\right),
\label{eq-tail-drift}
\end{eqnarray}
where we introduced the abbreviations $\overline{a}=(v^2+2)^{-t/2}/\Gamma(t)$
and $\overline{b}=\sqrt{v^2+2}-v\mathrm{sgn}(x)$. This non-Gaussian PDF may also
be obtained from application of the asymptotic behavior (\ref{bessel-approx}) of
the modified Bessel in Eq.~(\ref{pdf-flm-drift-exact}). Unlike the PDF of FLM
with external drift, which is symmetric and centered at $x=vt$, see
Eq.~(\ref{eq-tail-drift_2}), the non-Gaussian PDF of FLM with internal drift is
not symmetric when $v\neq0$, as shown in Fig.~\ref{fg-pdf-drift}(a); and when
$v=0$ it is reduced to Eq. ~\eqref{eq-pdf-flm-exponetial} .

The Laplace method can also be applied to obtain the approximation of the PDF in
the regime $|x|\ll t$, but this PDF cannot help us to capture the behavior of the
Gaussian domain of the PDF at long times, because the PDF is shifting with $vt$
as indicated by the first moment Eq.~(\ref{mu}). However, according to the
statistical properties of the process at long times, see above, we can reasonably
assume that the central portion of the PDF follows the Gaussian distribution
\begin{equation}
\label{eq-gau-drift}
P(x,t)\sim\frac{1}{\sqrt{2\pi(t^{2H}+v^2t)}}\exp\left(-\frac{(x-vt)^2}{2(t^{2H}
+v^2t)}\right).
\end{equation} 

In Fig.~\ref{fg-pdf-drift} we show stochastic simulations of the PDF for FLM
with internal drift ($v=1$) for the Hurst exponents $H=0.5$, $H=0.8$ and $H=0.2$.
Unlike the PDFs with external drift, the PDFs here are no longer symmetry,
although they still follow a Gaussian shape, Eq.~(\ref{eq-gau-drift}), in
certain regions, approximately given by $(vt-t^{H+1/2},vt+t^{H+1/2})$. For
$H=1/2$, the non-Gaussian behavior of the PDF far from the central part is
given by Eq.~(\ref{eq-tail-drift}).

\section{Discussion and Conclusions}
\label{sec5}

We investigate FLM, a generalized diffusion process obtained from subordination
of FBM to a gamma process, for the drift-free case and in the presence of both
an external and an internal drift $v$. The moments and MSDs of these two types
of drift show significantly different behaviors. While in both formulations the
second moment is ballistic at long times, the long time scaling of the MSDs are
$\simeq t^{2H}$ for external drift and a combination of $\simeq t^{2H}$ and
$\simeq t$ for internal drift. The latter thus shows apparently normal diffusion
$\simeq t$ at long times when $H<1/2$. Additionally, we examined the intricate
structure of the PDF and identify a central region, whose boundaries expand as
function of time, in which the PDF exhibits a Gaussian form. Outside of this
region, the PDF remains non-Gaussian. At long times, the non-Gaussian part
becomes increasingly suppressed by the central Gaussian region that the kurtosis
is no longer sensitive to the non-Gaussian tails. In particular, for an external
drift the PDF remains symmetric and centered at $x=vt$, whereas the internal
drift breaks this symmetry.

\begin{figure}
(a)\includegraphics[width=0.8\linewidth]{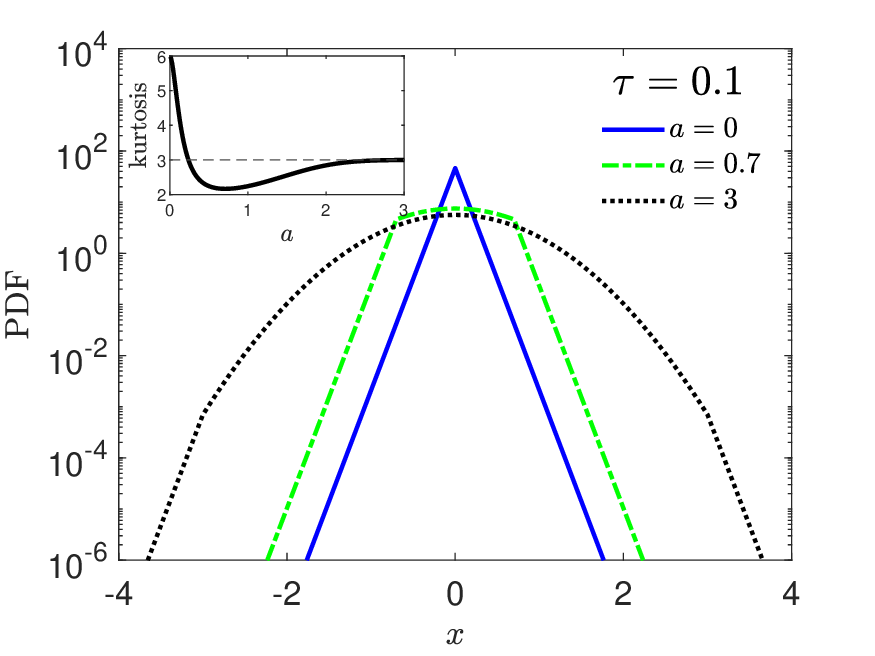}
(b)\includegraphics[width=0.8\linewidth]{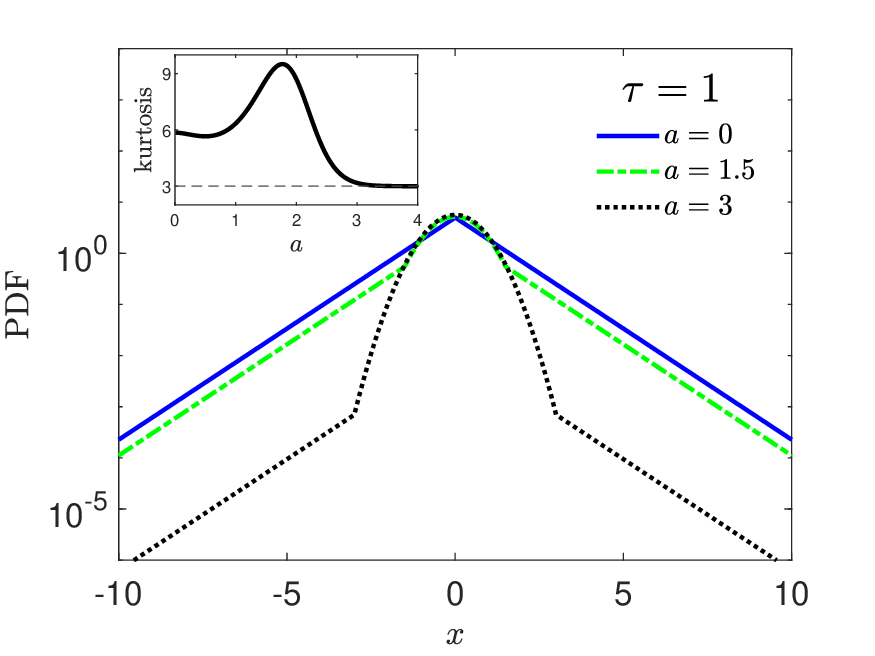}
(c)\includegraphics[width=0.8\linewidth]{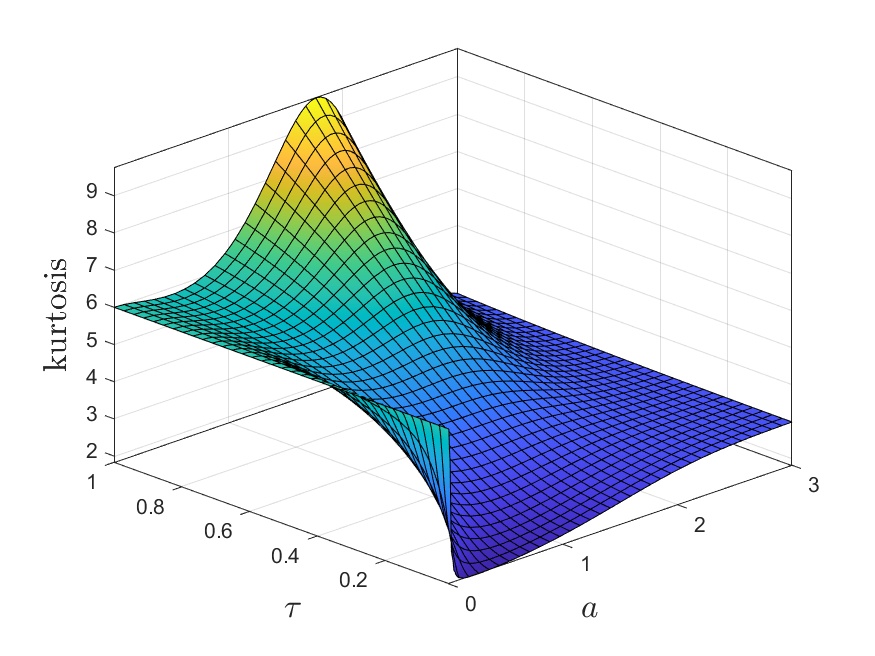}
\caption{PDF (\ref{combine}) with exponential tails for different crossover
position $a$ and scale parameters (a) $\tau=0.1$ and (b) $\tau=1$. The
corresponding kurtosis as function of $a$ is shown in the inset. (c) The
kurtosis as function of $a$ and $\tau$. The behavior of the kurtosis prior
to reaching the Gaussian limit value is influenced by the scale parameter
$\tau$.}
\label{exam-pdf}
\end{figure}

These interesting observations for the PDF can already be observed in the following
toy model for the composite, non-Gaussian PDF 
\begin{equation}
\label{combine}
f(x)=\mathscr{N}\left[e^{-a/\tau+a^2-x^2}\theta(a-|x|)+e^{-|x|/\tau}\theta(|x|
-a)\right],
\end{equation}
in which $\mathscr{N}$ is the normalization constant, $\theta(x)$ is the
Heaviside step function, and $\tau$ is a scale parameter of the exponential
distribution. The PDF (\ref{combine}) has a central Gaussian core for $|x|
\leq a$ and an exponential tail for $|x|>a$, as shown in Fig.~\ref{exam-pdf}.
The PDF quickly convergences to a Gaussian with increasing $a$, and its
kurtosis approaches the Gaussian value 3. Thus, despite the presence of the
non-Gaussian tails, the central Gaussian regime dominates the value of the
kurtosis when $a$ increases.

The key distinction between FLM and other comparable stochastic processes
such as FBM and L{\'e}vy motion lies in the fact that, in the latter two
cases, the PDFs of the increments maintain Gaussian and stable distributions
across all time scales. The PDF of FLM rather resembles that of the Diffusing
Diffusivity (DD) model \cite{chub14,chec17,jain16}, recently used to describe
Brownian yet non-Gaussian dynamics in various soft matter, biological, and
other complex systems \cite{wang12,wang09,miot21,rusc22,beta2024,beta2018,
schw2015,gold2009}. The DD model features a characteristic crossover timescale
from a non-Gaussian PDF at short times to a Gaussian PDF at long times. Our
findings suggest that FLM can serve as an alternative to the DD model for
capturing Brownian yet non-Gaussian behavior. However, it is worth noting
that the FLM and DD models can be distinguished by analyzing the behavior
of the kurtosis $\kappa$. In the FLM model, $\kappa$ approaches infinity
in the short-time limit, whereas in the DD model, a value of $\kappa=9$
(in the one-dimensional version) corresponding to a Laplace distribution
with additional power-law is observed.

Liang et al. \cite{metz23} recently investigated a subordinated fractional
Brownian motion (SFBM) with drift, establishing the PDF via subordination, in
a way analogous to Eq.~(\ref{eq-pdf-flm}). The DD model was also formulated
through a subordination approach \cite{chec17}, albeit using a different
function $h(s,t)$. The PDF of SFBM remains non-Gaussian throughout the entire
time range. Therefore, the transition behavior of the PDF is strongly linked
to the statistical properties of the subordination process $s(t)$ or the PDF
$h(s,t)$, which should be explored in future studies.

From an experimental perspective, the time-scale-dependent non-Gaussian
behavior plays a crucial role in FLM-type dynamics. In \cite{meer04,molz07},
FLM in the absence of a drift was proposed to examine the increments of
hydraulic conductivity data. Standard FLM was extended to accurately capture
the sampling time-scale-dependent PDF of bed elevation and sediment transport
rates in \cite{gant09}, where the use of operational time was justified by the
stochastic nature of turbulent velocity fluctuations near the bed, which
introduces randomness in the particle entrainment. With the inclusion of drift,
even richer behaviors are expected to emerge in complex systems across hydrology,
finance, soft- and bio-matter dynamics, as well as turbulence.

\begin{acknowledgments}

Y.L. acknowledges financial support from the Alexander von Humboldt Foundation
(Grant No. 1217531), the National Natural Science Foundation of China (Grant
No. 12372382) and the Qing Lan Project of Jiangsu Province (Grant No. 2024).
R.M. acknowledges financial support from the German Science Foundation (DFG,
Grant No. ME 1535/12-1) and NSF-BMBF CRCNS (Grant No. 2112862/STAXS). A.V.C.
acknowledges BMBF Project PLASMA-SPIN Energy (Grant No. 01DK21018).

\end{acknowledgments}

\appendix

\section{Asymptotic behavior of $K_\mu(z)$ for large order $\mu\gg1$}
\label{app-a}

The integral representation of the modified Bessel function of the third kind
with argument $z>0$ according to Ref.~\cite{abra72} reads
\begin{equation}
\label{A1}
K_\mu(z)=\frac{1}{2}\left(\frac{z}{2}\right)^\mu\int_0^{\infty}\exp\left(-s
-\frac{z^2}{4s}\right)\frac{ds}{s^{\mu+1}}.
\end{equation}
The asymptotic expansion of Eq.~(\ref{A1}) can be examined using the Laplace
method for large values of the order $\mu$.

First we rewrite Eq.~(\ref{A1}) in the form
\begin{equation}
K_\mu(z)=\frac{1}{2}\left(\frac{z}{2}\right)^\mu\int_0^{\infty}\exp\left(
-\Phi(s)\right)ds.
\end{equation}
with the function
\begin{equation}
\Phi(s)=s+\frac{z^2}{4s}+(\mu+1)\ln(s),
\end{equation}
which attains a simple quadratic minimum at
\begin{equation}
s_{\mathrm{min}}=\frac{-(\mu+1)+\sqrt{(\mu+1)^2+z^2}}{2}.
\end{equation}  

When $z\gg\mu\gg1$, we have the quadratic minimum
\begin{eqnarray}
s_{\mathrm{min}}\sim\frac{z}{2}
\end{eqnarray}
and use the standard Laplace method. The integral (\ref{A1}) can be estimated as
\begin{eqnarray}
\label{A6}
\nonumber
K_\mu(z)&\sim&\frac{1}{2}\left(\frac{z}{2}\right)^\mu\sqrt{\frac{2\pi}{|\Phi^{''}
(s_{\mathrm{min}})|}}\exp(-\Phi(s_{\mathrm{min}}))\\
&\sim&\sqrt{\frac{\pi}{2z}}e^{-z}.
\end{eqnarray}
We note that this approximation is consistent with the asymptotic behavior of
$K_\mu(z)$ for $z\to\infty$ given in Ref.~\cite{abra72}.

When $1\ll z\ll\mu$ the quadratic minimum asymptotically becomes
\begin{eqnarray}
s_{\mathrm{min}}\sim\frac{-\mu+\mu\sqrt{\left(1+z^2/{\mu^2}\right)}}{2}\sim
\frac{z^2}{4\mu}.    
\end{eqnarray}
Using the Laplace method, we find
\begin{eqnarray}
\nonumber
K_\mu(z)&\sim&\frac{2^{\mu-1}}{z^\mu}\left(\sqrt{2\pi}e^{-\mu}\mu^{\mu-1/2}
\right)\exp\left(-\frac{z^2}{4\mu}\right)\\
&\sim&\frac{\Gamma(\mu)}{2}\left(\frac{z}{2}\right)^{-\mu}\exp\left(-\frac{z^2}{
4\mu}\right).
\label{A8}
\end{eqnarray}
This approximation in the regime $z\ll\mu$, to our best knowledge, has not been
explored previously. Moreover, although the approximation in Eq.~(\ref{A8}) holds
for $1\ll z\ll\mu$, it is also consistent with $K_\mu(z)\sim[\Gamma(\mu)/2]\big(
z/\big)^{-\mu}$ as $z\to0$ in Ref.~\cite{abra72}.

\begin{figure}
\includegraphics[width=0.8\linewidth]{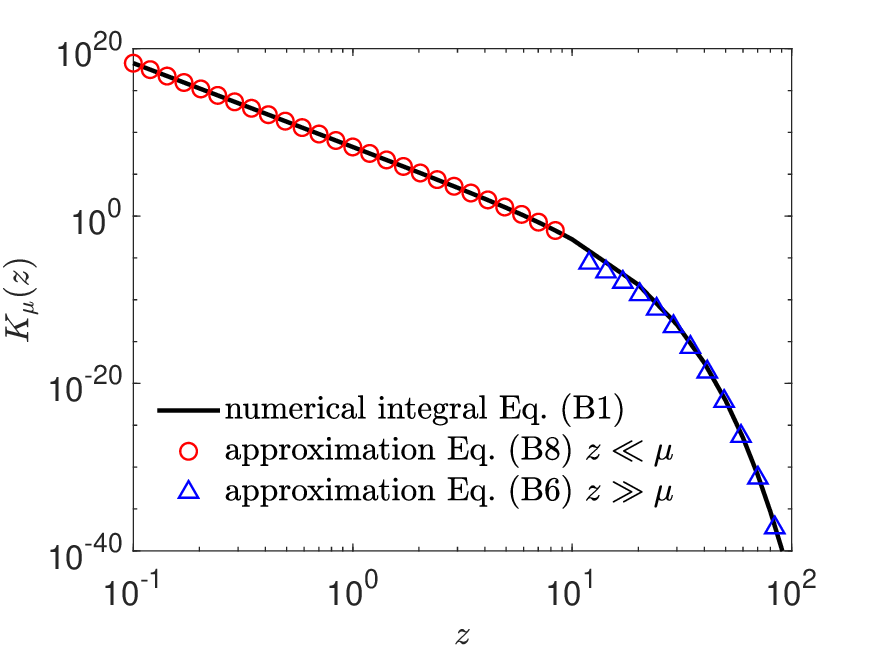}
\caption{Modified Bessel function $K_\mu(z)$, Eq.~(\ref{A1}), for large order
$\mu=10$. The blue triangles represent the approximate form (\ref{A6}) for
argument $z\gg\mu$, and the red circles show approximation (\ref{A8}) for $z\ll
\mu$.}
\label{S1}
\end{figure}

Figure \ref{S1} displays the approximate form of $K_\mu(z)$ for $z\gg\mu$ and
$z\ll\mu$, in a perfect agreement with the numerical integral (\ref{A1}).  

\section{Asymptotic behavior of $P(x,t)$ for $t\to\infty$ in
Eq.~(\ref{eq-pdf-flm-new})} 
\label{app-b}

We rewrite the PDF $P(x,t)$ in Eq.~(\ref{eq-pdf-flm-new}) in the form
\begin{equation}
\label{B1}
P(x,t)=\frac{1}{2H\sqrt{2\pi}\Gamma(t)}\mathcal{H}\left[\left(\frac{|x|}{
\sqrt{2}}\right)^{1/H}\right],
\end{equation} 
in which the $H-$function is expressed by the contour integral \cite{mathai}
\begin{eqnarray}
\mathcal{H}[z]&=&H_{0,2}^{2,0}\left[z\left|\begin{array}{l}\rule{1.2cm}{0.02cm}\\
(t-H,1),(0,1/(2H))\end{array}\right.\right]\\
&=&\frac{1}{2\pi i}\int\limits_L\Gamma(t-H+s)\Gamma\left(\frac{s}{2H}\right)
z^{-s}ds,
\label{B2}
\end{eqnarray}
where $L$ is a contour separating the poles of the two gamma functions. 

Then, using the property
\begin{equation}
\label{B3}
\Gamma(t-H+s)\sim\Gamma(t)t^{-H+s},\quad t\to\infty,
\end{equation}
of the gamma function [Eq.~(6.1.46) in \cite{abra72}], the $H-$function (\ref{B2})
can be approximated by  
\begin{equation}
\label{B4}
\mathcal{H}[z]\sim t^{-H}\Gamma(t)\times\frac{1}{2\pi i}\int\limits_L\Gamma
\left(\frac{s}{2H}\right)\left(\frac{z}{t}\right)^{-s}ds.
\end{equation}
With the variable transformation
\begin{equation}
\tau=\frac{s}{2H},\quad y=\left(\frac{z}{t}\right)^{2H}, 
\end{equation}
we find
\begin{eqnarray}
\nonumber
\mathcal{H}[z]&\sim&2Ht^{-H}\Gamma(t)\frac{1}{2\pi i}\int\limits_L\Gamma
\left(\tau\right)y^{-\tau}d\tau\\
&\sim& Ht^{-H}\Gamma(t)\exp\left(-\left[\frac{z}{t}\right]^{2H}\right).
\label{B6}
\end{eqnarray}
In Eq.~(\ref{B6}) we applied the inverse Mellin transform \cite{ober12}
\begin{equation}
\frac{1}{2\pi i}\int\limits_L\Gamma\left(\tau\right)y^{-\tau}d\tau=e^{-y}.
\end{equation}

Therefore substituting the $H-$function (\ref{B6}) with $z=\left(|x|/\sqrt{2}
\right)^{1/H}$ into the PDF (\ref{B1}), we arrive at the Gaussian approximation
of the PDF when $t\to\infty$, namely,
\begin{equation}
P(x,t)\sim\frac{1}{\sqrt{2\pi t^{2H}}}\exp\left(-\frac{x^2}{2t^{2H}}\right).
\end{equation}

\end{document}